\documentclass[aps,prl,reprint,amsmath,amssymb,showpacs,superscriptaddress]{revtex4-1}
\usepackage{CJK}
\usepackage{graphicx}% Include figure files
\usepackage{dcolumn}% Align table columns on decimal point
\usepackage{bm}% bold math
\usepackage{mathrsfs}   % added by S.B. Wang to use \mathscr{L}_{N\pi}
\usepackage{color,xcolor}  % added by S.B. Wang to use \color{red}
\usepackage{multirow}  % added by S.B. Wang to use multirow in table
\usepackage{booktabs} % added by S.B. Wang to use \cmidrule(r){5-6}
\usepackage{mathtools}  % added by S.B. Wang to use prescript

\usepackage{hyperref}% add hypertext capabilities
\newcommand{\bbm}{\begin{bmatrix}}
\newcommand{\ebm}{\end{bmatrix}}
\newcommand{\bBm}{\begin{Bmatrix}}
\newcommand{\eBm}{\end{Bmatrix}}
\newcommand{\bpm}{\begin{pmatrix}}
\newcommand{\epm}{\end{pmatrix}}

\usepackage{amsfonts,amssymb,stmaryrd,latexsym,amsmath,braket}
\usepackage{graphicx,subfigure}
\usepackage{comment}
\usepackage{times}
\usepackage{slashed}
\usepackage{bm}
\usepackage{braket}
\usepackage{chapterbib}
\usepackage{color}

\usepackage[export]{adjustbox}
\usepackage{multirow}
\usepackage{makecell}
\usepackage{booktabs}
\usepackage{array}
\newcommand{\PreserveBackslash}[1]{\let\temp=\\#1\let\\=\temp}
\newcolumntype{C}[1]{>{\PreserveBackslash\centering}p{#1}}
\newcolumntype{R}[1]{>{\PreserveBackslash\raggedleft}p{#1}}
\newcolumntype{L}[1]{>{\PreserveBackslash\raggedright}p{#1}}

\newcommand{\beginsupplement}{%
        \setcounter{table}{0}
        \renewcommand{\thetable}{S\arabic{table}}%
        \setcounter{figure}{0}
        \renewcommand{\thefigure}{S\arabic{figure}}%
        \setcounter{equation}{0}
        \renewcommand{\theequation}{S\arabic{equation}}%
     }

\makeatletter
\let\saved@includegraphics\includegraphics
\AtBeginDocument{\let\includegraphics\saved@includegraphics}
\makeatother

\begin{document}
% \begin{CJK*}{UTF8}{song} % Use default fonts from CJK (see below)

% Use the \preprint command to place your local institutional report
% number in the upper righthand corner of the title page in preprint mode.
% Multiple \preprint commands are allowed.
% Use the 'preprintnumbers' class option to override journal defaults
% to display numbers if necessary

%Title of paper
\title{{\em Ab initio} calculation of hyper-neutron matter}

\author{Hui~Tong}
%\email{}
\affiliation{Helmholtz-Institut f\"{u}r Strahlen- und Kernphysik and Bethe Center for Theoretical Physics, Universit\"{a}t Bonn, D-53115 Bonn, Germany}

\author{Serdar Elhatisari}
%\email{}
\affiliation{Faculty of Natural Sciences and Engineering, Gaziantep Islam Science and Technology University, Gaziantep 27010, Turkey}
\affiliation{Interdisciplinary Research Center for Industrial Nuclear Energy (IRC-INE), King Fahd University of Petroleum and Minerals (KFUPM), 31261 Dhahran, Saudi Arabia}
\affiliation{Helmholtz-Institut f\"{u}r Strahlen- und Kernphysik and Bethe Center for Theoretical Physics, Universit\"{a}t Bonn, D-53115 Bonn, Germany}

\author{Ulf-G.~Mei{\ss}ner}
%\email{}
\affiliation{Helmholtz-Institut f\"{u}r Strahlen- und Kernphysik and Bethe Center for Theoretical Physics, Universit\"{a}t Bonn, D-53115 Bonn, Germany}
\affiliation{Institute~for~Advanced~Simulation~(IAS-4),~Forschungszentrum~J\"{u}lich,~D-52425~J\"{u}lich,~Germany}
\affiliation{Center for Advanced Simulation and Analytics (CASA),~Forschungszentrum~J\"{u}lich,~D-52425~J\"{u}lich,~Germany}
\affiliation{Tbilisi State University, 0186 Tbilisi, Georgia}
\affiliation{Peng Huanwu Collaborative Center for Research and Education,  Beihang University, Beijing 100191, China}

\date{\today}

% \begin{abstract}
% \end{abstract}

% insert suggested PACS numbers in braces on next line
%\pacs{26.60.-c,21.65.Ef,21.60.Jz,21.60.De}

%21.60.Jz Nuclear Density Functional Theory and extensions
%21.10.Gv Nucleon distributions and halo features
%21.30.-x Nuclear forces
%21.10.-k Properties of nuclei; nuclear energy levels
%21.10.Re Collective levels
%21.60.Ev Collective models
%21.60.Cs Shell model
%21.45.Ff Three-nucleon forces
%23.20.-g Electromagnetic transitions
%23.20.Js Multipole matrix element
%26.60.-c Nuclear matter aspects of neutron stars
%21.65.Ef Symmetry energy
%21.60.-n Nuclear structure models and methods
%27.20.+n  6 A 19
%27.60.+j 90  A 149
%27.50.+e 59  A  89
%21.60.De Ab initio methods
% insert suggested keywords - APS authors don't need to do this
%\keywords{}

%\maketitle must follow title, authors, abstract, \pacs, and \keywords
\maketitle
%\end{CJK*}

% body of paper here - Use proper section commands
% References should be done using the \cite, \ref, and \label commands

In the era of multi-messenger astronomy, neutron stars arguably stand out as the most captivating astrophysical objects~\cite{LIGOScientific:2017vwq1}.
Neutron stars consist of the densest form of baryonic matter observed in the universe, and within their interiors, exotic new forms of matter may exist~\cite{Lattimer:2004pg1}.
With the detection of various neutron star phenomena in recent years,  such as gravitational waves and electromagnetic radiation, more valuable information regarding the mysterious dense matter within their cores will be unraveled.
These findings, together with the measurements of the masses or radii,  strongly constrain the neutron star matter equation of state (EoS) and theoretical models of their composition.
However, the observation of neutron star masses above $2.0M_\odot$ has ruled out many predictions of exotic non-nucleonic components. Resolving this problem, known as the hyperon puzzle, is crucial for understanding the complex interplay between strong nuclear forces and the behavior of dense matter under extreme conditions~\cite{Schulze:2011zza1,Lonardoni:2014bwa1}.

In this study, we use the framework of Nuclear Lattice Effective Field Theory (NLEFT)~\cite{Lahde:2019npb1} to gain new insights into the generation of hyperons, more specifically $\Lambda(1116)$ particles, within dense environments.
To enable calculations with arbitrary numbers of nucleons and hyperons using only one auxiliary field, we introduce a novel formulation of the auxiliary field quantum Monte Carlo (AFQMC) algorithm, which allows for more accurate and efficient simulations free of sign oscillations.
Additionally, we incorporate two-body $N\Lambda$ and $\Lambda \Lambda$ interactions, as well as three-body terms such as $NN\Lambda$ and $N\Lambda\Lambda$, based on the minimal nuclear interaction model~\cite{Lu:2018bat1}, into the pionless effective field theory for nucleons.
Initially, we focus on systems consisting solely of nucleons and determine the low-energy constants parameterizing the $2N$ and the $3N$ forces by constraining them to the saturation properties of symmetric nuclear matter, as it is well-known that fixing the $3N$ forces in light nuclei leads to a serious overbinding in heavier systems~\cite{Elhatisari:2022zrb1} if mostly local forces are employed.
After constructing our interactions, we perform predictive calculations for the EoS of pure neutron matter (PNM) by considering up to 232 neutrons in a box to achieve densities up to five times the empirical saturation density of nuclear matter, i.e., $\rho =0.8$ fm$^{-3}$.
Our results for the EoS of PNM are in very good agreement with \emph{ab initio} calculations using chiral interactions up to N3LO~\cite{Elhatisari:2022zrb1} within the given density range.
Subsequently, we introduce $\Lambda$-particles into our framework and determine the parameters of the $N\Lambda$ and $\Lambda\Lambda$ interactions by fitting them to experimental data, including the $N\Lambda$ cross section~\cite{Sechi-Zorn:1968mao1} and the $\Lambda\Lambda$ ${}^1S_0$ scattering phase shift from chiral effective field theory~\cite{Haidenbauer:2015zqb1}, respectively.
%\textbf{
The $NN\Lambda$ and $N \Lambda \Lambda$ forces are further constrained by the separation energies of single- and double-$\Lambda$ hypernuclei, spanning systems from ${}_\Lambda^5$He to ${}_{\Lambda\Lambda}^{~~6}$Be, denoted as HNM(I).
It is difficult to  probe the behavior of the EoS at high densities encountered in neutron stars in terrestrial laboratories, and various phenomenological schemes~\cite{Weber:1989uq1} and microscopical models~\cite{Schulze:2011zza1} suggest that hyperons emerge in the inner core of neutron stars at densities around $\rho \approx (2-3)\rho_0$.
Therefore, similar to using the saturation properties of symmetric nuclear matter to pin down the three-nucleon forces (3NFs), we alternatively determined the $NN\Lambda$ and $N \Lambda \Lambda$ forces by using the separation energies of hypernuclei and the $\Lambda$ threshold densities $\rho_\Lambda^{\rm th}$ around $(2-3)\rho_0$ simultaneously in HNM(II) and HNM(III).
We set $\rho_\Lambda^{\rm th}$ = 0.398(2)(5) and 0.520(2)(6)~fm$^{-3}$ for HNM(II) and HNM(III), respectively.
In the next step, we perform simulations for hyper-neutron matter by including up to 116 hyperons in the box and calculate the corresponding EoSs.
More details on the construction of the actions underlying PNM EoS and the three variants of HNM are given in the Supplementary materials.

The results for pure neutron matter and hyper-neutron matter are presented from our state-of-art nuclear lattice simulations. HNM is composed of neutrons and $\Lambda$ hyperons, where $\rho_N$, $\rho_\Lambda$, and $\rho=\rho_N+\rho_\Lambda$ are the neutron, $\Lambda$ hyperon and total baryon density of the system, respectively, and $x_\Lambda=\rho_\Lambda/\rho$ is the fraction of $\Lambda$ hyperons. The $\Lambda$ threshold densities $\rho_\Lambda^{\rm th}$ is determined by imposing the equilibrium condition $\mu_N=\mu_\Lambda$, where the chemical potentials for neutrons $\mu_N$ and lambdas $\mu_\Lambda$ are evaluated via the derivatives of the energy density $\varepsilon_{\scriptscriptstyle \rm HNM}$,
\begin{equation}
  \mu_N(\rho,x_\Lambda)=\frac{\partial \varepsilon_{\scriptscriptstyle \rm HNM}}{\partial \rho_N}, ~~\mu_\Lambda(\rho,x_\Lambda)=\frac{\partial \varepsilon_{\scriptscriptstyle \rm HNM}}{\partial \rho_\Lambda},
\end{equation}
which indicates that an
accurate determination of the chemical potentials necessitates computing the energy density for various densities and different numbers of $\Lambda$ hyperons.

\begin{figure}[htbp]
  \centering
  \includegraphics[height=6.3cm]{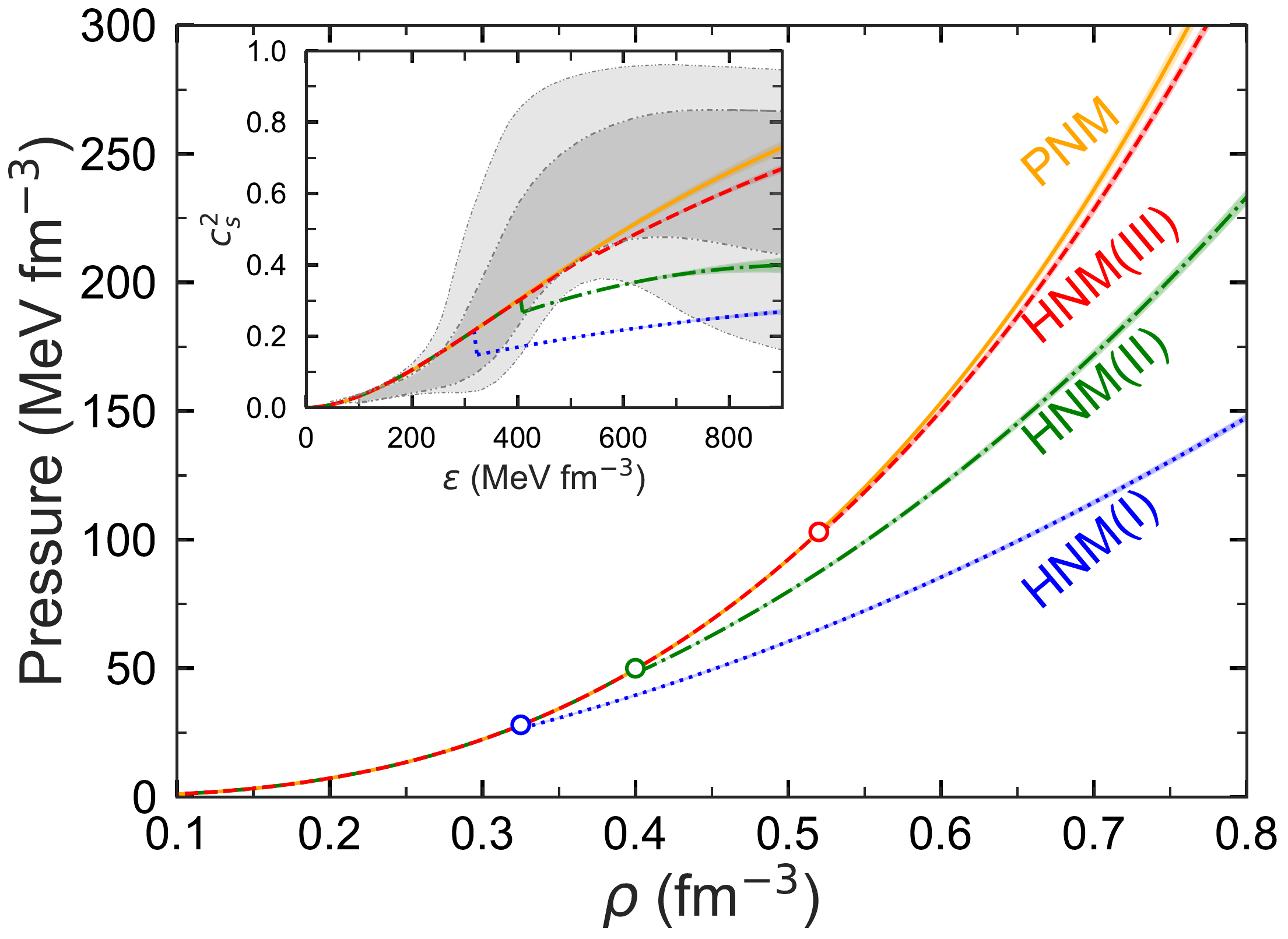}
  \caption{(Color online) EoS for HNM. The orange solid curve denotes pure neutron matter, obtained from the $NN$ and $NNN$ interactions. The red dashed line represents the EoS of HNM with hyperons interacting via the two-body interactions ($N\Lambda$ and $\Lambda\Lambda$) and the third set of three-body hyperon-nucleon interaction ($NN\Lambda$ and $N\Lambda\Lambda$). The blue dotted curve and the green dot-dashed curve are calculated with the first and second sets of three-body hyperon-nucleon interactions. The $\Lambda$ threshold densities $\rho_\Lambda^{\rm th}$ are marked by open circles. In the inset, the speed of sound corresponding to the PNM and HNM EOSs is shown. The gray shaded regions are the inference of the speed of sound for neutron star matter in view of the recent observational data~\cite{Brandes:2023hma1}.}
  \label{fig1}
\end{figure}

In Fig.~\ref{fig1}, the EoSs for PNM and for HNM are displayed.
As anticipated, the inclusion of hyperons results in a softer EoS, and the threshold density is $\rho_\Lambda^{\rm th} = 0.325(2)(4)$~fm$^{-3}$ for HNM(I).
Here and what follows, the first (second) error is the statistical (systematic) one.
This threshold aligns with predictions from various phenomenological schemes~\cite{Weber:1989uq1} and microscopical models~\cite{Schulze:2011zza1}, which suggest that hyperons emerge in the inner core of neutron stars at densities around $\rho \approx (2-3)\rho_0$.
Furthermore, we construct two additional variants of HNM, denoted as HNM(II) and HNM(III).
%These variants are designed to have threshold densities within the predicted region, with HNM(II) and HNM(III) threshold densities increasing to 0.398(2)(5) fm$^{-3}$ and 0.520(2)(6) fm$^{-3}$, respectively.
The EoS becomes stiffer at higher densities for these variants, indicating the inclusion of more repulsion in the three-body hyperon-nucleon interactions.
The squared speed of sound, $c_s^2$, is also shown in the inset of Fig.~\ref{fig1}.
It is observed that the causality limit ($c_s^2<1$) is fulfilled for both PNM and HNM.
It should be noted that in the pioneering calculations of Lonardoni et al.~\cite{Lonardoni:2014bwa1}, they performed auxiliary field diffusion Monte Carlo (AFDMC) simulations with $N_n = 38, 54, 66$ neutrons and their PNM EoS is stiffer than ours and exceeds the causality limit for the speed of sound at densities above $\rho \simeq 0.68$~fm$^{-3}$.
The EoS characterized by nucleonic degrees of freedom exclusively demonstrate a monotonic increase in $c_s^2$ with increasing energy density.
The appearances of $\Lambda$ hyperons, however, induces changes in this behavior, leading to non-monotonic curves that signify the incorporation of additional degrees of freedom.
The onset of $\Lambda$ hyperons precipitates a sharp reduction in the speed of sound, marking a significant transition in the stiffness of the EoS.
For comparison, the constraints on $c_s^2$ within the interiors of neutron stars inferred by a Bayesian inference method are also shown~\cite{Brandes:2023hma1}.
These constraints are established based on recent multi-messenger data, in combination with limiting conditions from nuclear physics at low densities, as depicted by the gray shaded regions.
%See Ref.~\cite{Koehn:2024set} for a review and Ref.~\cite{Essick:2021ezp} for a detailed analysis using employed recent astronomical data.
The results for PNM and HNM(III) agree well with the marginal posterior probability distributions at the 95\% and 68\% levels.
%Note that one should consider $\beta$-stable nuclear matter for a more realistic description of neutron stars, since the proton fraction may reach around 20\%-30\% at the center of neutron stars (but is much smaller otherwise)~\cite{Tong:2022yml,Bombaci:2018ksa}.
It should be noted that we used the minimal nuclear interaction model, not the potential models derived from chiral effective field theory~\cite{Elhatisari:2022zrb1}. In Ref.~\cite{Elhatisari:2022zrb1}, a full chiral interactions at next-to-next-to-next-to-leading order (N3LO) was used for PNM and symmetric nuclear matter. This significantly increases the computational cost so that the calculations were done only at densities up to $2\rho_0$. In the future, we will incorporate the methodological advancements introduced in this work to explore neutron star EoS calculations using higher order chiral forces at densities larger than $2\rho_0$. While our calculations extend into the higher-density regime, we recognize that the behavior of EoS at these densities is less constrained, and the nuclear interaction models we employ may introduce uncertainties. To validity the nuclear interaction model in this work, we have quantified the theoretical uncertainty due to the six different three-nucleon forces which are shown in Fig.~\ref{fig1} and Fig.~S2. We find the theoretical uncertainty in Fig.~S2 is significantly smaller than the empirical uncertainty of the nuclear matter, and the uncertainty is also quite small for PNM in Fig.~\ref{fig1}. This uncertainty quantification provides a solid foundation for future research on uncertainties in hyper-neutron matter, as the hyperonic three-baryon interactions used in our calculations follow the same uncertainty quantification approach. Therefore, the uncertainty in our nuclear interaction models has been significantly reduced and is well controlled.
Due to the current limitations of computational resources, calculating the EoS for arbitrary fractions of protons, neutrons, and other hyperons is still very challenging in any lattice approach.
This can be achieved when more computational resources become available in the future and it will allow us to avoid the errors introduced by using the so-called symmetry energy approximation.
%In addition, the presence of other hyperons, such as $\Sigma$ and $\Xi$, will be influenced by hyperonic two-body and three-body forces in lattice calculations. This can be further studied once more experimental data on these hypernuclei become available.
%Note, however, the neutron stars in general have a small proton fraction, which is neglected in the present work.

\begin{figure}[htbp]
  \centering
  \includegraphics[height=6.3cm]{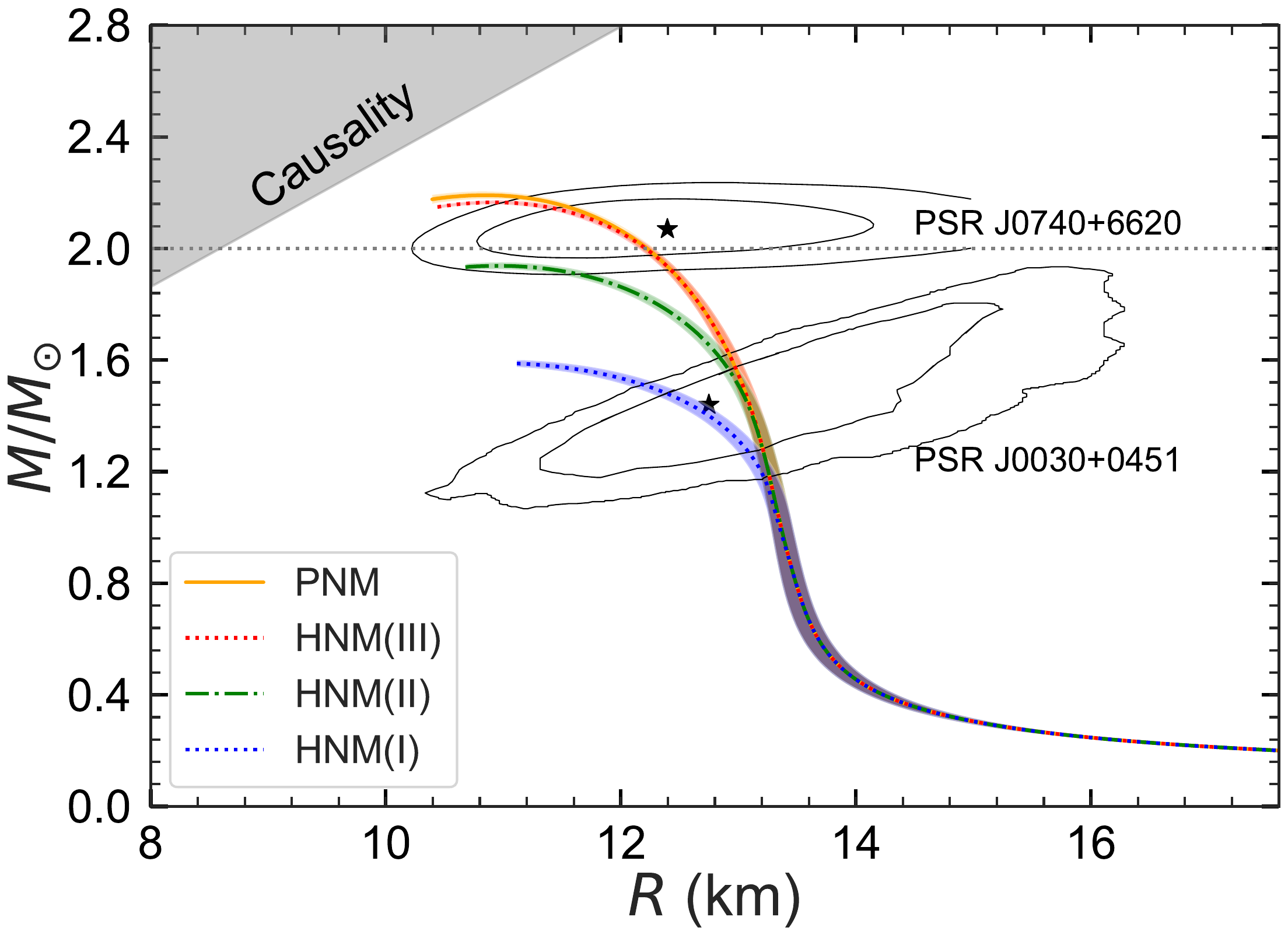}
  \caption{(Color online) Neutron star mass-radius relation. The legend is the same as of Fig.~\ref{fig1}. The gray horizontal dotted line represents 2$M_\odot$. The inner and outer contours indicate the allowed area of mass and radius of neutron stars by NICER’s analysis of PSR~J0030+0451~\cite{Miller:2019cac1} and PSR~J0740+6620~\cite{Riley:2021pdl1}. The excluded causality region is also shown by the grey shaded region~\cite{Lattimer:2006xb1}.}
  \label{fig2}
\end{figure}

The ``holy grail'' of neutron-star structure, the mass-radius (MR) relation, is displayed in Fig.~\ref{fig2}.
These relations for PNM and HNM are obtained by solving the Tolman-Oppenheimer-Volkoff (TOV) equations with the EoSs of Fig.~\ref{fig2}.
The appearance of $\Lambda$ hyperons in neutron star matter remarkably reduces the predicted maximum mass compared to the PNM scenario.
The maximum mass for PNM, HNM(I), HNM(II), and HNM(III) are 2.19(1)(1)~$M_\odot$, 1.59(1)(1)~$M_\odot$, 1.94(1)(1)~$M_\odot$, and 2.17(1)(1)~$M_\odot$, respectively.
Three neutron stars have been measured to have gravitational masses close to $2M_\odot$: PSR J1614-2230, with $M = 1.908 \pm 0.016~M_\odot$~\cite{Demorest:2010bx1}; PSR J0348+0432, with $2.01 \pm 0.04~M_\odot$~\cite{Antoniadis:2013pzd1}; and PSR J0740+6620, with $2.08 \pm 0.07~M_\odot$~\cite{Fonseca:2021wxt1}.
These measurements significantly constrain the EoS of dense nuclear matter, ruling out the majority of currently proposed EoSs with hyperons from phenomenological approaches~\cite{Burgio:2021vgk1}.
Our results show that the inclusion of the $NN\Lambda$ and $N\Lambda\Lambda$ interaction in HNM(III) leads to an EoS stiff enough such that the resulting neutron star maximum mass is compatible with the three mentioned measurements of neutron star masses.
Therefore, the repulsion introduced by the hyperonic three-body interactions plays a crucial role, since it substantially increases the value of the $\Lambda$ threshold density. It is also noteworthy that HNM(I) predicts a maximum mass above the canonical neutron mass of 1.4$M_\odot$, whereas the model~(I) incorporating repulsive $NN\Lambda$ interactions in the auxiliary field diffusion Monte Carlo~\cite{Lonardoni:2014bwa1}, Hartree-Fock~\cite{Djapo:2008au1}, and Brueckner-Hartree-Fock (BHF)~\cite{Schulze:2011zza1,Wang:2024jtl1} calculations yield values below 1.4$M_\odot$.
%In the multimessenger era, another important constraint of the canonical neutron star mass (1.4$M_\odot$) is the tidal deformability $\Lambda_{1.4M_\odot}$ and radius $R_{1.4M_\odot}$.
%The tidal deformability $\Lambda_{1.4M_\odot}$ for PNM, HNM(I), HNM(II), and HNM(III) are 597(5)(18), 451(5)(31), 587(5)(19), and 597(5)(18), respectively, see also Fig.~S4 in~\cite{SM}.
%The initial estimation for the tidal deformability $\Lambda_{1.4M_\odot}$ has an upper bound $\Lambda_{1.4M_\odot}<800$~\cite{LIGOScientific:2017vwq} from the observation of BNS merger event GW170817. Then a revised analysis from the LIGO and Virgo collaborations gave $\Lambda_{1.4M_\odot}=190_{-120}^{+390}$~\cite{LIGOScientific:2018cki}.
%It is important to underscore that our results are located in these regions and agree well with the one inferred in Ref.~\cite{Fasano:2019zwm} for the two neutron stars in the merger event GW170817 at the 90\% level.
In addition, the radii corresponding to PNM, HNM(I), HNM(II), and HNM(III) are $R_{1.4M\odot}=13.10(1)(7)$~km, $R_{1.4M\odot}=12.71(4)(13)$~km, $R_{1.4M\odot}=13.09(1)(8)$~km, and $R_{1.4M\odot}=13.10(1)(7)$~km, in order.
Our results for the neutron star radii are also consistent with the constraints by NICER~\cite{Miller:2019cac1} for the mass and radius of PSR J0030+0451, i.e., mass $1.44_{-0.14}^{+0.15}M_\odot$ with radius $13.02_{-1.06}^{+1.24}$~km.
The 68\% and 95\% contours of the joint probability density distribution of the mass and radius from the NICER analysis are also shown in Fig.~\ref{fig2}.
%Our results for the neutron star radii are also consistent with those of other works, such as $R_{1.4M_\odot}\leqslant 13.76$~km~\cite{Fattoyev:2017jql}, $R_{1.4M_\odot}\leqslant 13.6$~km~\cite{Annala:2017llu}, and $12.00$~km $\leqslant R_{1.4M_\odot}\leqslant 13.45$~km~\cite{Most:2018hfd} from the tidal deformability~\cite{LIGOScientific:2017vwq,LIGOScientific:2018hze},  $9.7$~km $\leqslant R_{1.4M_\odot}\leqslant 13.9$~km from the chiral effective field theory with the constraint $M = 1.97~M_\odot$~\cite{Hebeler:2013nza}, and the constraints by NICER~\cite{Miller:2019cac} for the mass and radius of PSR J0030+0451, i.e., mass $1.44_{-0.14}^{+0.15}M_\odot$ with radius $13.02_{-1.06}^{+1.24}$~km. The 68\% and 95\% contours of the joint probability density distribution of the mass and radius from the NICER analysis are also shown in Fig.~\ref{fig3}.
We further note that despite the significant reduction in the fraction of $\Lambda$ hyperons caused by the hyperonic three-body force in HNM(III), they still exist within the interior of a $2.17M_\odot$ neutron star.
%, this was also observed by the BHF calculations with the chiral hyperonic three-body forces~\cite{Logoteta:2019utx}, see also Fig.~S7 in~\cite{SM}.
This is different from the conclusion drawn in Ref.~\cite{Lonardoni:2014bwa1}, where it was found the hyperonic three-body force in their parametrization (II)  capable of generating an EoS stiff enough to support maximum masses consistent with the observations of $2M_\odot$ neutron stars results in the complete absence of $\Lambda$ hyperons in the cores of these objects. We also note that while model HNM(III) successfully supports the mass of PSR J0740+6620 with $2.08 \pm 0.07~M_\odot$, the $M(R)$ curve appears only marginally compatible with the combined experimental data. This issue is primarily associated with two factors. First, the characteristics of this neutron star mass region are predominantly determined by the EoS at higher densities. The baryon-baryon interactions considered in this work account only for contributions from the minimal interaction model, whereas higher-order baryon-baryon interactions are expected to influence the EoS at higher densities. Therefore, it will be interesting and necessary to include the higher-order baryon-baryon interactions in the subsequent work. Second, the rotation of a neutron star induces corresponding increases in both its mass and radius, which can lead to a better agreement with the combined experimental data. The rotational frequency of this neutron star is 346 Hz, while the current calculations have been limited to static case. In the next step, we will incorporate the properties of rotating neutron stars.

In summary, we have performed the first lattice Monte Carlo calculation of hyper-neutron matter with a large number of neutrons and $\Lambda$s and derived the resulting properties of neutron stars. In the next steps, one should include the proton fraction, other hyperons of the baryon octet, and make use of the recently developed high-fidelity chiral interactions at N3LO~\cite{Elhatisari:2022zrb1}, though this will pose a formidable computational challenge.

{\bf Conflict of interest}

The authors declare that they have no conflict of interest.

{\bf  Acknowledgments}

We are grateful for discussions with members and partners of the Nuclear Lattice Effective Field Theory Collaboration, in particular Zhengxue Ren. We are deeply thankful to Wolfram Weise for some thoughtful comments.
Hui Tong thanks Jie Meng and Sibo Wang for helpful discussions. Serdar Elhatisari thanks Dean Lee for useful discussions on the auxiliary field formulations. We acknowledge funding by the European Research Council (ERC) under the European Union's
Horizon 2020 research and innovation programme (AdG EXOTIC, grant agreement No. 101018170), by the MKW
NRW under the funding code NW21-024-A and by DFG and NSFC
through funds provided to the Sino-German CRC 110 “Symmetries and the Emergence of Structure in
QCD” (NSFC Grant No. 12070131001, DFG Project-ID 196253076)).
The work of Serdar Elhatisari was further supported by  the Scientific and Technological Research Council of Turkey (TUBITAK project no. 120F341).
The work of Ulf-G.~Mei{\ss}ner was further supported by CAS through the President's International Fellowship Initiative (PIFI)
(Grant No. 2025PD0022).

{\bf Author contributions}

The project was initiated and supervised by Ulf-G.~Mei{\ss}ner. Serdar Elhatisari conceived the AFMC formulation for hypernuclear system, and code development, testing and optimization were led by Serdar Elhatisari with contributions by Hui Tong. Hui Tong performed the analysis of the data with contributions by Serdar Elhatisari, led production runs, and created the figures. All authors contributed to the writing.

%{\bf Appendix A. Supplementary materials}

%Supplementary materials to this short communication can befound online at.

% \bibliography{TF-NM}

\clearpage

\beginsupplement
\onecolumngrid

\section{Supplemental Materials}

\subsection{Nuclear Lattice Effective Field Theory}
\subsubsection{Lattice Formalism}
\label{sec:Lattice-Hamiltonian}

Lattice effective field theory is a quantum many-body method that synthesises the theoretical framework of effective field theory (EFT) with powerful numerical approaches~\cite{Lee:2008fa,Lahde:2019npb}. The method has
been applied to describe the properties of atomic nuclei~\cite{Borasoy:2005yc} and neutron matter~\cite{Lee:2004qd} in pionless EFT at leading order (LO), and to perform the first \textit{ab initio} calculation of the  Hoyle state in the spectrum of $^{12}$C~\cite{Epelbaum:2011md} and
$\alpha$-$\alpha$ scattering~\cite{Elhatisari:2015iga} in chiral
EFT at next-to-next-to-leading order (N2LO). Moreover, it has recently been applied to compute the properties of atomic nuclei and the equation of state of neutron and symmetric nuclear matter in chiral
EFT at next-to-next-to-next-to-leading order (N3LO)~\cite{Elhatisari:2022zrb}. In addition, the method has been used in formulating an EFT with only four parameters and built on Wigner's SU(4) spin-isospin symmetry~\cite{Wigner:1936dx}. This EFT effectively captures gross properties of light and medium-mass nuclei and the equation of state of neutron matter with remarkable accuracy, typically within a few percent~\cite{Lu:2018bat}. Noteworthy applications of this EFT include the study of the first \textit{ab
initio} thermodynamics calculation of nuclear clustering~\cite{Lu:2019nbg} and microscopic investigations of clusters in hot dilute matter using the method of light-cluster distillation~\cite{Ren:2023ued}, and the identification of the emergent geometry and intrinsic cluster structure of the low-lying states of $^{12}$C~\cite{Shen:2021kqr,Shen:2022bak}. Additionally, it has been utilized in resolving  the puzzle of the alpha-particle monopole transition form factor~\cite{Meissner:2023cvo}.

Building upon the significant achievements of the EFT within Wigner's SU(4) spin-isospin symmetry, which we refer to as the minimal nuclear interaction, throughout this paper we exclusively define and employ pionless EFT at LO for nucleons (see also~\cite{Konig:2016utl}), derived from this minimal nuclear interaction. This approach allows us to make use of the well-established theoretical framework by the minimal nuclear interaction, providing a solid basis for our calculations for hyper-neutron matter equations of state. It is important to note that our calculations consider only $\Lambda$ hyperons, with the inclusion of $\Sigma$ hyperons reserved for future work. Note that the $\Lambda$-$\Sigma^0$ transition induces three-body forces which are effectively
represented by $\Lambda NN$ forces here.

For the hyperon-nucleon and hyperon-hyperon interactions, we also utilize minimal interactions assuming that these interactions are spin symmetric. Therefore, the Hamiltonian is defined as,
\begin{align}
H= & H_{\rm free}+\frac{c_{NN}}{2}\sum_{\vec{n}}
\,:\,\left[
\tilde{\rho}(\vec{n})
\right]^2
\,:\,
+\frac{c_{NN}^{T}}{2}\sum_{I,\vec{n}}
\,:\,
\left[
\tilde{\rho}_{I}(\vec{n})
\right]^2
\,:\,
\nonumber\\
& + c_{N\Lambda}\sum_{\vec{n}}
\,:\,
\tilde{\rho}(\vec{n})
\tilde{\xi}(\vec{n})
\,:\,
+ \frac{c_{\Lambda\Lambda}}{2}\sum_{\vec{n}}
\,:\,
\left[
\tilde{\xi}(\vec{n})
\right]^2
\,:\,
\nonumber\\
 &
+V^{\rm GIR}_{NN}
+V^{\rm GIR}_{N\Lambda}
+V^{\rm GIR}_{\Lambda\Lambda}
+V_{\rm Coulomb}
\nonumber\\
 &
+V_{NNN}
+V_{NN\Lambda}
+V_{N\Lambda\Lambda}
\,,
\label{eq:H-001}
\end{align}
where $H_{\rm free}$ is the kinetic energy term defined by using fast Fourier transforms to produce the exact dispersion relations $E_N = p^2/(2m_{N})$ and $E_\Lambda =p^2/(2m_{\Lambda})$ with nucleon mass $m_{N}=938.92$~MeV and hyperon mass $m_{\Lambda}=1115.68$~MeV, the $::$ symbol indicates normal ordering, $c_{NN}$ is the coupling constant of the SU(4) symmetric short-range two-nucleon interaction, $c_{NN}^{T}$ is the coupling constant of the isospin-dependent short-range two-nucleon interaction, that breaks SU(4) symmetry (see the discussion below), $c_{N\Lambda}$ ($c_{\Lambda\Lambda}$) is the coupling constant of the spin-symmetric short-ranged hyperon-nucleon (hyperon-hyperon) interaction, and $\tilde{\rho}$ ($\tilde{\xi}$) is nucleon (hyperon) density operator, that is smeared both locally and non-locally,
\begin{align}
\tilde{\rho}(\vec{n}) = \sum_{i,j=0,1}
\tilde{a}^{\dagger}_{i,j}(\vec{n}) \, \tilde{a}^{\,}_{i,j}(\vec{n})
+
s_{\rm L}
 \sum_{|\vec{n}-\vec{n}^{\prime}|^2 = 1}
 \,
 \sum_{i,j=0,1}
\tilde{a}^{\dagger}_{i,j}(\vec{n}^{\prime}) \, \tilde{a}^{\,}_{i,j}(\vec{n}^{\prime})
\,,
\end{align}
\begin{align}
\tilde{\rho}_{I}(\vec{n})= \sum_{i,j,j^{\prime}=0,1}
\tilde{a}^{\dagger}_{i,j}(\vec{n}) \,\left[\tau_{I}\right]_{j,j^{\prime}} \, \tilde{a}^{\,}_{i,j^{\prime}}(\vec{n})
+
s_{\rm L}
 \sum_{|\vec{n}-\vec{n}^{\prime}|^2 = 1}
 \,
  \sum_{i,j,j^{\prime}=0,1}
\tilde{a}^{\dagger}_{i,j}(\vec{n}^{\prime}) \,\left[\tau_{I}\right]_{j,j^{\prime}} \, \tilde{a}^{\,}_{i,j^{\prime}}(\vec{n}^{\prime})\,,
\end{align}
\begin{align}
\tilde{\xi}(\vec{n}) = \sum_{i=0,1}
\tilde{b}^{\dagger}_{i}(\vec{n}) \, \tilde{b}^{\,}_{i}(\vec{n})
+
s_{\rm L}
 \sum_{|\vec{n}-\vec{n}^{\prime}|^2 = 1}
 \,
 \sum_{i=0,1}
\tilde{b}^{\dagger}_{i}(\vec{n}^{\prime}) \, \tilde{b}^{\,}_{i}(\vec{n}^{\prime})
\,.
\end{align}
   The smeared annihilation and creation operators, $\tilde{a}$ ($\tilde{b}$) and $\tilde{a}^{\dagger}$ ($\tilde{b}^{\dagger}$) for nucleons (hyperons), have with spin $i = 0, 1$ (up, down) and isospin $j = 0, 1$ (proton, neutron) indices,
\begin{align}
\tilde{a}_{i,j}(\vec{n})=a_{i,j}(\vec{n})+s_{\rm NL}\sum_{|\vec{n}^{\prime}-\vec{n}|=1}a_{i,j}(\vec{n}^{\prime})\,,
\label{eqn:rho-local-nonlocal}
\end{align}
\begin{align}
\tilde{b}_{i}(\vec{n})=b_{i}(\vec{n})+s_{\rm NL}\sum_{|\vec{n}^{\prime}-\vec{n}|=1}b_{i}(\vec{n}^{\prime}).
\label{eqn:xi-local-nonlocal}
\end{align}
In Eq.~(\ref{eq:H-001}), $V_{\text{Coulomb}}$ represents the Coulomb interaction, and for the details we direct the reader to Ref.~\cite{Li:2018ymw}. The nonlocal smearing applied on the lattice introduces an explicit dependence on the center-of-mass momentum, thereby breaking Galilean invariance. Consequently, in Eq.~(\ref{eq:H-001}) we introduce $V^{\text{GIR}}_{NN}$, $V^{\text{GIR}}_{N\Lambda}$, and $V^{\text{GIR}}_{\Lambda\Lambda}$, which denote the Galilean invariance restoration (GIR) interactions for the nucleon-nucleon, nucleon-hyperon, and hyperon-hyperon interactions, respectively. We refer the reader to Ref.~\cite{Li:2019ldq} for further details.

Finally, we introduce the three-baryon interactions $V_{NNN}$, $V_{NN\Lambda}$, and $V_{N\Lambda\Lambda}$, given in Eq.~(\ref{eq:H-001}). Recent \emph{ab-initio} nuclear structure and scattering calculations have revealed the significant impact of locally smeared interactions on nuclear binding \cite{Elhatisari:2016owd}. Hence, the three-baryon interactions utilized in our calculations are defined with two different choices of local smearing,
\begin{align}\label{eq:VNNN}
V_{NNN}
=
\sum_{i = 1,2}
\frac{c_{NNN}^{(d_i)}}{6}
\,
\sum_{\vec{n}}
\,:\,
\left[
\rho^{(d_i)}(\vec{n})
\right]^3
\,:\,,
\end{align}
where the parameter $d_i$ denotes the range of local smearing with $0 \leq d_1 < d_2 \leq 3$ (in lattice units). Similarly, the three-baryon interactions consisting of two nucleons and one hyperon are defined with two different choices of local smearing,
\begin{align}
V_{NN\Lambda}
=
\sum_{i = 1,2}
\frac{c_{NN\Lambda}^{(d_i)}}{2}
\,
\sum_{\vec{n}}
\,:\,
\left[
\rho^{(d_i)}(\vec{n})
\right]^2 \xi^{(d_i)}(\vec{n})
\,:\,,
\label{eqn:V-NNY}
\end{align}
%Similarly, the three-baryon interaction consisting of two nucleons and one hyperon is defined by one specific choice of local smearing,
%\begin{align}
%V_{NN\Lambda}
%=
%\frac{c_{NN\Lambda}}{2}
%\,
%\sum_{\vec{n}}
%\,:\,
%\left[
%\rho^{(1)}(\vec{n})
%\right]^2 \xi^{(1)}(\vec{n})
%\,:\,
%\,,
%\end{align}
and the interactions involving one nucleon and two hyperons are expressed by also two different choices of local smearing,
\begin{align}
V_{N\Lambda\Lambda}
=
\sum_{i = 1,2}
\frac{c_{N\Lambda\Lambda}^{(d_i)}}{2}
\,
\sum_{\vec{n}} \,
 \,:\,
 \rho^{(d_i)}(\vec{n})  \,
\left[
\xi^{(d_i)}(\vec{n})
\right]^2
\,:\,,
\label{eqn:V-NYY}
\end{align}
%and the interaction involving one nucleon and two hyperons is expressed by also one specific choice of local smearing,
%\begin{align}
%V_{N\Lambda\Lambda}
%=
%\frac{c_{N\Lambda\Lambda}}{2}
%\,
%\sum_{\vec{n}} \,
% \,:\,
% \rho^{(1)}(\vec{n})  \,
%\left[
%\xi^{(1)}(\vec{n})
%\right]^2
%\,:\,
%\,,
%\end{align}
where ${\rho}$ (${\xi}$) is then purely locally smeared nucleon (hyperon) density operator with annihilation and creation operators, ${a}$ (${b}$) and ${a}^{\dagger}$ (${b}^{\dagger}$) for nucleons (hyperons),
\begin{align}
{\rho}^{(d)}(\vec{n}) = \sum_{i,j=0,1}
{a}^{\dagger}_{i,j}(\vec{n}) \, {a}^{\,}_{i,j}(\vec{n})
+
s^{\rm 3B}_{\rm L}
 \sum_{|\vec{n}-\vec{n}^{\prime}|^2 = 1}^{d}
 \,
 \sum_{i,j=0,1}
{a}^{\dagger}_{i,j}(\vec{n}^{\prime}) \, {a}^{\,}_{i,j}(\vec{n}^{\prime})
\,,
\label{eqn:rho-local}
\end{align}
\begin{align}
{\xi}^{(d)}(\vec{n}) = \sum_{i=0,1}
{b}^{\dagger}_{i}(\vec{n}) \, {b}^{\,}_{i}(\vec{n})
+
s^{\rm 3B}_{\rm L}
 \sum_{|\vec{n}-\vec{n}^{\prime}|^2 = 1}^{d}
 \,
 \sum_{i=0,1}
{b}^{\dagger}_{i}(\vec{n}^{\prime}) \, {b}^{\,}_{i}(\vec{n}^{\prime})
\,.
\label{eqn:xi-local}
\end{align}
Here, the parameter $d$ gives the range of local smearing as pointed out above, and $s^{\rm 3B}_{\rm L}$ defines the strength of the local smearing. In our analysis of locally smeared three-baryon forces given in the above equations, we exclusively consider smearing with ranges $d \leq 3$ in lattice units, corresponding to a physical distance of $1.9$~fm. In addition, in Eqs.~(\ref{eqn:rho-local}) and (\ref{eqn:xi-local}) the local smearing refers to interactions that do not change the positions of particles, while in Eqs.~(\ref{eqn:rho-local-nonlocal}) and (\ref{eqn:xi-local-nonlocal}) the nonlocal smearing specifies interactions that do change the relative positions of particles.
The numerical values of the various LECs and lattice parameters are given below when we discuss nuclei, symmetric nuclear matter as well
as hyper-nuclei. We note that throughout we assume that the appearance of the Fermi momentum $k_F$
as a new scale in the problem does not require a re-ordering of the interaction terms, see
e.g. Ref.~\cite{Meissner:2001gz}.

%\begin{figure}[htbp]
%  \centering\includegraphics[height=7.0 cm,valign=t]{figs/figS5.pdf}
%  %\hspace{0.5cm}
%  %\includegraphics[height=5.1 cm,valign=t]{figs/}
%  \caption{Nucleon-Nucleon scattering phase shift.
%  %{\bf Left Panel:
%  %{\bf Right Panel:}
%  \label{figS5}}
%\end{figure}

%\begin{table}[htbp]
%    \centering
%    \caption{Summary of interaction parameters.}
%    \begin{tabular}{|l|c|c|c|c|c|c|c|}
%    \hline
%                 & $c_{NN}$ & $s_L$ & $s_{NL}$ & GIR0 & GIR1 & GIR2 & cutoff  \\
%    \hline
%    $^1$S$_0$  & $-1.21368\times 10^{-7}$ & 0.06 & 0.6 & $-1.74359\times 10^{-7}$ & $4.16751\times 10^{-8}$ & $-6.30765\times 10^{-9}$ & 180 \\
%    $^3$S$_1$  & $-1.91717\times 10^{-7}$ & 0.06 & 0.6 & $-2.91812\times 10^{-7}$ & $7.03009\times 10^{-8}$ & $-1.08328\times 10^{-8}$ & 180 \\
%    \hline
%    \end{tabular}
%\end{table}

\subsubsection{Auxiliary Field Formulation for Hypernuclear Systems}

For a first attempt to investigate $\Lambda N$ scattering on the lattice, we refer to~\cite{ShahinMSc}.
The incorporation of the $\Lambda$ into the nuclear lattice effective field theory framework was considered in Ref.~\cite{Frame:2020mvv} using the impurity lattice Monte Carlo (ILMC) method~\cite{Elhatisari:2014lka}, and this study involved calculating the binding energies of light hypernuclei ${}_{\Lambda}^{3}$H, ${}_{\Lambda}^{4}$H, and ${}_{\Lambda}^{5}$He. The ILMC method treats minority species of fermions, such as hyperons in a nucleus, as worldlines in a medium of majority species of particles simulated by the Auxiliary Field Quantum Monte Carlo (AFQMC) method. Recently, the ILMC method has been extended to enable the study of systems with two impurities~\cite{Hildenbrand:2022imw}. In the present work, we propose a novel approach that allows for the efficient investigation of hypernuclear systems with an arbitrary number of hyperons.

In our lattice simulations, we employ the AFQMC method as it leads to a significant suppression of the sign oscillations. For a comprehensive overview of lattice simulations, the reader is directed to Ref.~\cite{Lahde:2019npb}. AFQMC represents a powerful computational framework within quantum many-body physics, particularly tailored for investigating strongly correlated systems. This method addresses the challenge of solving the full $A$-body Schr\"odinger equation by introducing a Hubbard-Stratonovich transformation. This transformation incorporates auxiliary fields to decouple particle densities, thereby enhancing the applicability of Monte Carlo techniques. In essence, within the AFQMC formalism, individual nucleons evolve as if they are single particles in a fluctuating background of auxiliary fields.

The following discussion begins with a discrete auxiliary field formulation for the SU(4) symmetric short ranged two-nucleon interaction given in Eq.~(\ref{eq:H-001}),
\begin{align}
   : \exp \left( -\frac{a_{t} \, c_{NN}}{2} \, \tilde{\rho}^2
          \right) :
=
\sum_{k = 1}^{3} \, w_{k} \,
: \exp \left( \sqrt{-a_{t} \, c_{NN}}  \, s_{k} \, \tilde{\rho} \right) \, :
\label{eqn:AFQMC-NN}
\end{align}
where $a_{t}$ is the temporal lattice spacing. From a Taylor expansion of Eq.~(\ref{eqn:AFQMC-NN}) we determine the constants $s_{k}$ and $w_k$ as $s_{1} = -s_{3}=\sqrt{3}$, $s_{2} = 0$, $w_{1} = w_3 = 1/6$ and $w_2 = 2/3$.

The nucleon-nucleon interaction given in Eq.~(\ref{eqn:AFQMC-NN}) obeys the Wigner SU(4) symmetry~\cite{Wigner:1936dx}, which arises from the realization that the combined spin ($S$) and isospin ($T$) degrees of freedom of nucleons can be described by a single unified symmetry group. Since we use minimal forces for the hyperon-nucleon and hyperon-hyperon interactions, we now aim to derive an auxiliary field formulation for systems including neutrons, protons and $\Lambda$ hyperons. This derivation involves replacing the isospin SU$_T(2)$ with flavor SU$_F(3)$ within Wigner's SU(4) symmetry framework, and the combined spin ($S$) and flavor ($F$) invariance ultimately leads to the SU(6) symmetry~\cite{Gursey:1964htz}. However, the fact that the strengths of the nucleon-nucleon and hyperon-nuclear interactions are different is breaking this SU(6) symmetry, and there is no longer an approximate symmetry similar to Wigner's SU(4) symmetry used in Eq.~(\ref{eqn:AFQMC-NN}). Nevertheless, in the following we exploit the fact that $|c_{NN}|>|c_{N\Lambda}|> |c_{\Lambda\Lambda}|$, which is allowing us to introduce an auxiliary field formulation with an approximate SU(6) symmetry that protects our simulations including $\Lambda$ hyperons against strong sign oscillations.

The spin and isospin independent two-baryon interactions in Eq.~(\ref{eq:H-001}) is expressed as,
\begin{align}
V_{\rm 2B} = \frac{c_{NN}}{2}\sum_{\vec{n}}
\,:\,
\left[
\tilde{\rho}(\vec{n})
\right]^2
\,:\,
+ c_{N\Lambda}\sum_{\vec{n}}
\,:\,
\tilde{\rho}(\vec{n})
\tilde{\xi}(\vec{n})
\,:\,
+ \frac{c_{\Lambda\Lambda}}{2}\sum_{\vec{n}}
\,:\,
\left[
\tilde{\xi}(\vec{n})
\right]^2
\,:\,
\,,
\label{eqn:NY-Potential-000}
\end{align}
and this potential (\ref{eqn:NY-Potential-000}) can be rewritten in the following form,
\begin{align}
V_{\rm 2B} = \frac{c_{NN}}{2}\sum_{\vec{n}}
\,:\,
\left[
\tilde{\slashed{\rho}}(\vec{n})
\right]^2
\,:\,
+
\frac{1}{2}
\left(
c_{\Lambda\Lambda}
-\frac{c_{N\Lambda}^2}{c_{NN}}
\right)
\sum_{\vec{n}}
\,:\,
\left[
\tilde{\xi}(\vec{n})
\right]^2
\,:\,
\,,
\label{eqn:NY-Potential-010}
\end{align}
where $\tilde{\slashed{\rho}}$ is defined as,
\begin{align}
\tilde{\slashed{\rho}} =
\tilde{\rho} +
\frac{c_{ N\Lambda}}{c_{NN}} \,  \tilde{\xi}
\,.
\label{eqn:Density-rho-bar}
\end{align}
In Eq.~(\ref{eqn:NY-Potential-010}) the leading contribution comes from the first term in the right-hand side and it is treated non-perturbatively, while the remaining term is computed using first-order perturbation theory. Hence, we define a new Hubbard-Stratonovich transformation for the first term in Eq.~(\ref{eqn:NY-Potential-010}), enabling the simulations of systems consisting of both arbitrary number of nucleons and  arbitrary number of $\Lambda$ hyperons with a single auxiliary field,
\begin{align}
   : \exp \left( -\frac{a_{t} \, c_{NN}}{2} \, \tilde{\slashed{\rho}}^2\right) :
=
\sum_{k = 1}^{3} \, w_{k} \,
: \exp \left( \sqrt{-a_{t} \, c_{NN}}  \, s_{k} \,  \tilde{\slashed{\rho}} \right) \,. :
\label{eqn:AFQMC-NY}
\end{align}
It is evident that the solution for the auxiliary field variables $s_{k}$ and weights $w_{k}$ is consistent with systems containing only nucleons.

The AFQMC method introduced here broadens hypernuclear calculations by enabling simulations with any number of hyperons. In addition, the approach can be effectively applied to wide range of systems~\cite{Sedrakian:2005zj,PhysRevLett.112.195301}. Let us consider two distinct family of particles and call them $A$ and $B$, and assume that all interactions are attractive. When the square of the interaction strength between particle types $A$ and $B$, denoted as $c_{AB}^2$, is of comparable magnitude to the product of the interaction strengths within the same particle types, $c_{AA}c_{BB}$, the overall coupling of the second term in Eq.~(\ref{eqn:NY-Potential-010}) becomes very small, enabling perturbative treatment and calculations with a single auxiliary field. Furthermore, when $c_{AA} c_{BB} \gg c_{AB}^2$, the second term's overall coupling is attractive, calculations still can be performed with two auxiliary fields. However, only in the case of $c_{AA} c_{BB} \ll c_{AB}^2$, the overall coupling of the second term becomes repulsive which leads to significant sign problems.

Finally, we discuss the two-nucleon interaction $\sim c_{NN}^T$,
known to break SU(4) symmetry and to induce significant sign oscillations,
which was previously disregarded in minimal nuclear interaction studies~\cite{Lu:2018bat,Lu:2019nbg,Ren:2023ued,Shen:2021kqr,Shen:2022bak,Meissner:2023cvo}.
In this work, aimed at constraining nuclear forces by using the ground state energies of finite hypernuclei and the saturation properties of symmetric nuclear matter,
this isospin interaction is treated non-perturbatively. We employ a Hubbard-Stratonovich transformation and introduce a discrete auxiliary field defined as,
\begin{align}
    : \exp \left( -\frac{a_{t} \, c_{NN}^T}{2} \, \sum_{I} \, \tilde{\rho}_{I}^2
           \right) :
 =
 \sum_{k = 1}^{3} \, w_{k} \,
 : \exp \left( \sqrt{-a_{t} \, c_{NN}^{T}}  \, \sum_{I} \,  s_{k,I} \, \tilde{\rho}_{I} \right) \, :
 \label{eqn:AFQMC-NN-I}
 \end{align}
 To minimize the occuring sign oscillations in finite nuclei, we focus on systems with equal numbers of protons and neutrons. Furthermore, in the simulations
 of pure neutron matter and hyper-neutron matter, this term can be omitted due to the absence of particles breaking isospin symmetry,
 allowing for sign oscillation-free simulations.

\subsubsection{Lattice and computational details}

Throughout our calculations presented here, we use a spatial lattice spacing of $a = 1.1$~fm and a temporal lattice spacing of $a_t = 0.2$~fm. We use the local smearing parameter $s_{\rm L}=0.06$ and nonlocal smearing parameter $s_{\rm NL} = 0.6$, influencing the strength of locality and nonlocality of the two-baryon interactions, respectively. For the three-baryon interaction, we set the local smearing parameter to $s^{\rm 3B}_{\rm L} = 0.06$. To compute the ground state energies of finite nuclei and hypernuclei, we utilize various periodic cubic lattices ranging in length from $13.2$~fm to $19.7$~fm. We perform our calculations at different finite Euclidean time steps and extrapolate to the infinite Euclidean time limit using a single and double exponential  ansatz~\cite{Lahde:2019npb}. Furthermore, for the computation of pure neutron matter and hyper-neutron matter energies we use lattices with a length of $6.6$~fm and impose the average twisted boundary conditions (ATBC) to efficiently eliminate finite volume effects. For further details on ATBC and the extensive analysis demonstrating the negligible impact of finite volume effects when employing ATBC, we refer the reader to Ref.~\cite{Li:2019ldq}.

%---------------------------------------------------
\subsection{Symmetric nuclear matter}
%---------------------------------------------------

\begin{figure}[htbp]
  \includegraphics[width=0.5\textwidth]{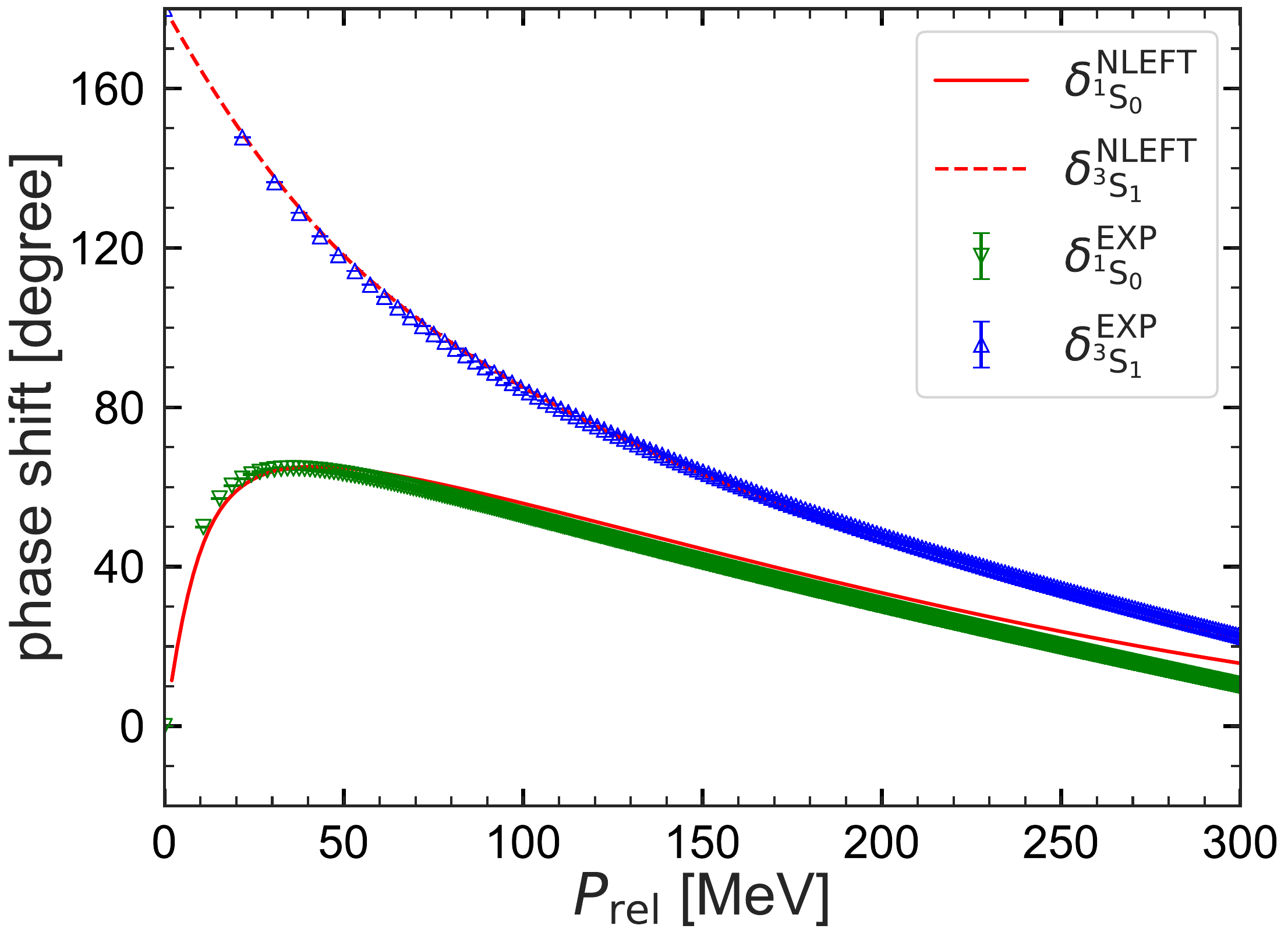}
  %\centering\includegraphics[width=0.5\textwidth,valign=t]{figS5.pdf}
  %\hspace{0.5cm}
  %\includegraphics[height=5.1 cm,valign=t]{figs/}
  \caption{Plots of the neutron-proton scattering phase shifts as functions of relative momenta. The blue up-triangles and green down-triangles denote the ${}^3{\rm S}_1$ and ${}^1{\rm S}_0$ phase shifts extracted from the Nijmegen partial wave analysis~\cite{Stoks:1993tb}, respectively. The solid and dash-dotted curve represents the lattice results for the ${}^1{\rm S}_0$  and the ${}^3{\rm S}_1$ phase shift, in order.
  \label{figS5}}
\end{figure}

Before considering the effect of hyperons on the neutron matter EoS, we determine the unknown LECs of the two- and three-nucleon interactions given in Eq.~(\ref{eq:H-001}) and predict the EoS corresponding to PNM. For PNM, the neutron-neutron interaction is dominated by the attractive component of the $^1S_0$ partial wave, and the $(T=0)$ $^3S_1-^3D_1$ state does not contribute to the $(T = 1)$ neutron-neutron state. The effects of higher partial waves, like $P$-wave, will be considered in our future work, though the contributions from the $P$-wave are expected to partially offset each other~\cite{HeYeTongLang:2013ihv}. In the first step, we pin down the nucleon-nucleon interactions by fitting to the two $S$-wave phase shifts of nucleon-nucleon scattering as shown in Fig.~\ref{figS5}. From these independent scattering phase shift fits we determine the coupling constants as $c_{{}^1{\rm S}_0}=-1.21\times10^{-7}$~MeV$^{-2}$ and $c_{{}^3{\rm S}_1}=-1.92\times10^{-7}$~MeV$^{-2}$ corresponding with the spin-singlet isospin-triplet and the spin-triplet isospin-singlet channel, respectively, which are related to the LECs given in Eq.~\eqref{eq:H-001} via
\begin{equation}
 c_{NN}^{} = (3\,c_{{}^1{\rm S}_0} + c_{{}^3{\rm S}_1})/4, ~~
c_{NN}^{T} = (c_{{}^1{\rm S}_0} - c_{{}^3{\rm S}_1})/4.
\end{equation}

\begin{comment}
\begin{table}[htbp]
    \centering
    \caption{Summary of the NN interaction parameters.}
    \begin{tabular}{|l|c|c|c|c|c|c|c|}
    \hline
                 & $c_{NN}$ & $s_L$ & $s_{NL}$ & $c_{GIR,0}$ & $c_{GIR,1}$ & $c_{GIR,2}$ & $a$~[fm]  \\
    \hline
    $^1$S$_0$  & $-1.21\times 10^{-7}$ & 0.06 & 0.6 & $-1.74\times 10^{-7}$ & $4.17\times 10^{-8}$ & $-6.31\times 10^{-9}$ & 1.1 \\
    $^3$S$_1$  & $-1.92\times 10^{-7}$ & 0.06 & 0.6 & $-2.92\times 10^{-7}$ & $7.03\times 10^{-8}$ & $-1.08\times 10^{-8}$ & 1.1 \\
    \hline
    \end{tabular}
    \label{tabNN}
\end{table}
\end{comment}
\begin{figure}[htbp]
  \includegraphics[width=0.5\textwidth]{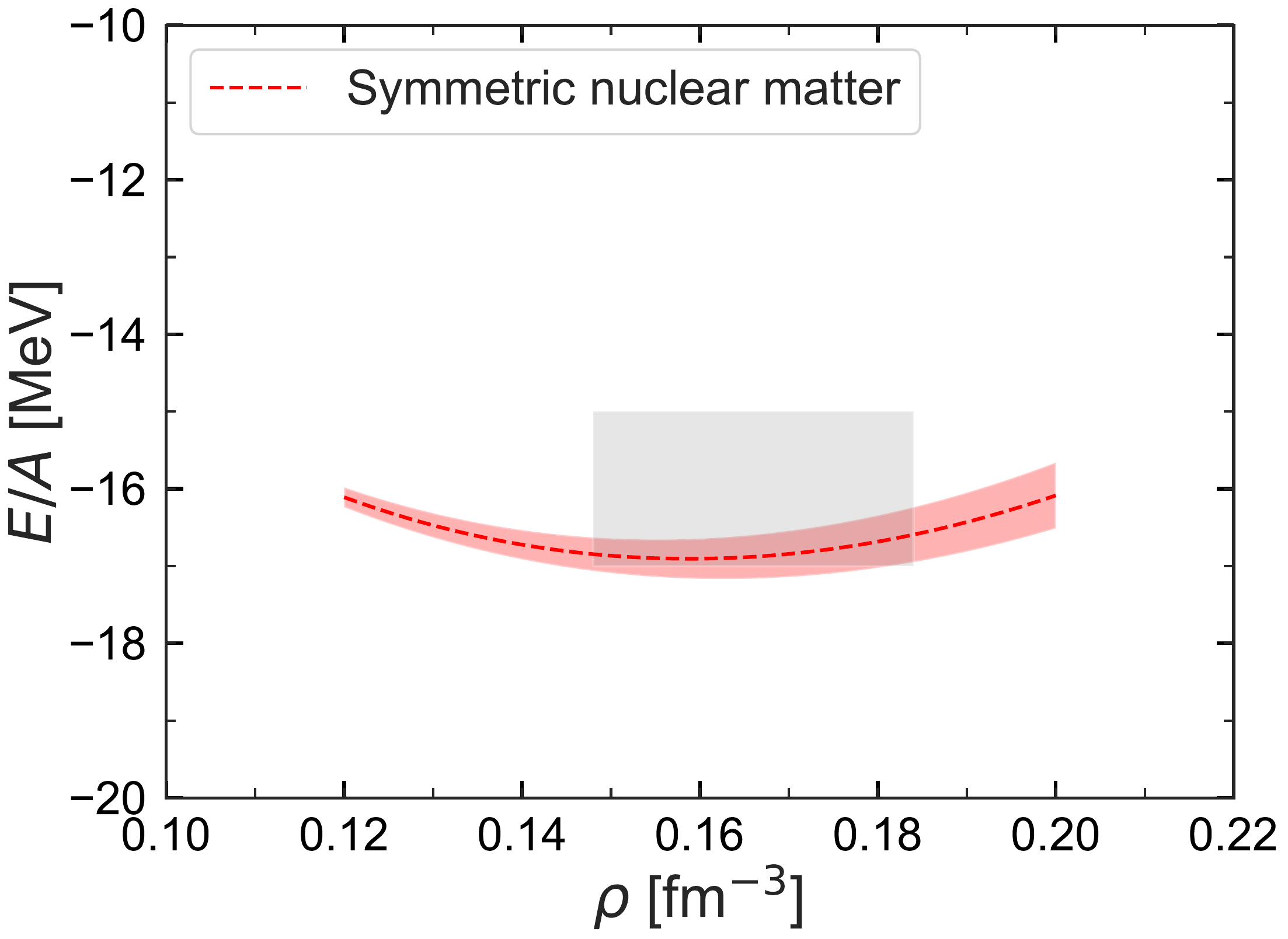}
  %\centering\includegraphics[width=0.5\textwidth,valign=t]{figS1.pdf}
  %\hspace{0.5cm}
  %\includegraphics[height=5.1 cm,valign=t]{figs/}
  \caption{Energy per nucleon as a function of density for symmetric nuclear matter from the NLEFT calculations. The gray shaded area indicates the empirical values.
  %{\bf Left Panel:
  %{\bf Right Panel:}
  \label{figS1}}
\end{figure}

In the next step, we determine the two LECs of the three-nucleon forces given in Eq.~\eqref{eq:VNNN}. This is accomplished by obtaining best fits to the saturation properties of symmetric nuclear matter considering all possible combinations of $d_1$ and $d_2$ with $0 \leq d_1 < d_2 \leq 3$. Through this process, we arrive at six different sets of interactions, enabling us to quantify the theoretical uncertainty of our calculations. The results for the energy per nucleon in symmetric nuclear matter from the NLEFT calculations are illustrated in Fig.~\ref{figS1}. The red shaded area represents the theoretical uncertainty in energies resulting from six different sets of three-nucleon interactions, while the red dashed line denotes the mean value for the energy per nucleon in symmetric nuclear matter, corresponding to the $3N$ interactions with LECs $c_{NNN}^{(1)}=4.98\times10^{-12}$~MeV$^{-5}$ and $c_{NNN}^{(3)}=1.80\times10^{-12}$~MeV$^{-5}$. The gray shaded area denotes the empirical values.
%\textbf{
We find the theoretical uncertainty in Fig.~\ref{figS1} is significantly smaller than the empirical uncertainty of the nuclear matter.
%}
As a prediction, we find for the compression modulus $K_\infty = 229.0(3.6)$~MeV, in good agreement with the
empirical value of $K_\infty = 240(20)$~MeV \cite{Garg:2018uam}. The symmetry energy $E_{\rm sym}$ at the saturation density is $35.3\pm{0.6}$~MeV, which agree with the ranges of (30-35)~MeV~based on chiral EFT interactions~\cite{Hebeler:2009iv,Tews:2012fj,Holt:2016pjb,Drischler:2017wtt,Jiang:2020the,Keller:2022crb,Lim:2023dbk}, see also
the review~\cite{Oertel:2016bki}.

%---------------------------------------------------
\subsection{Hypernuclei}
%---------------------------------------------------

\begin{figure}[tp]
  \includegraphics[height=6.20 cm,valign=t]{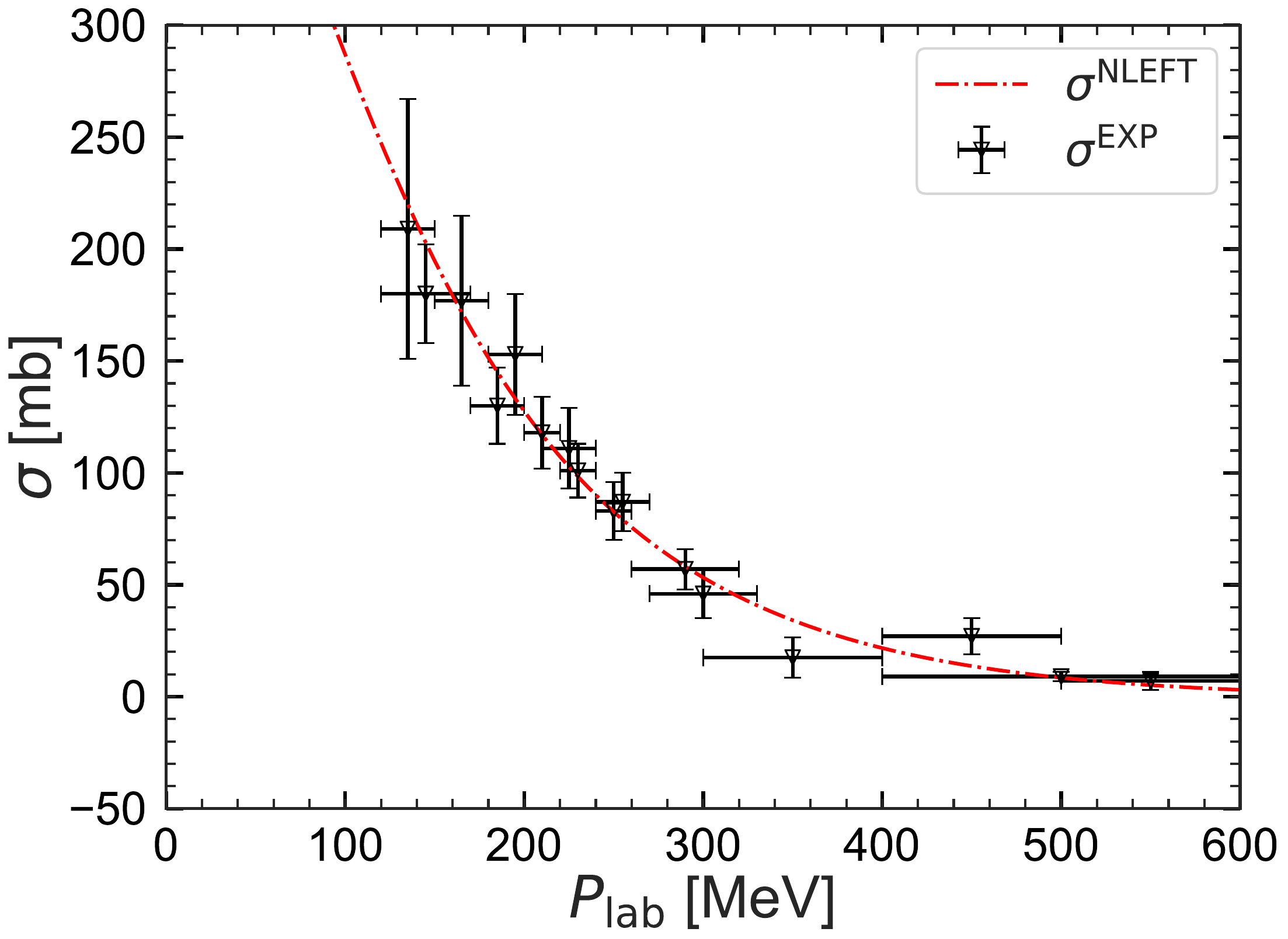}~~~~
  \includegraphics[height=6.20 cm,valign=t]{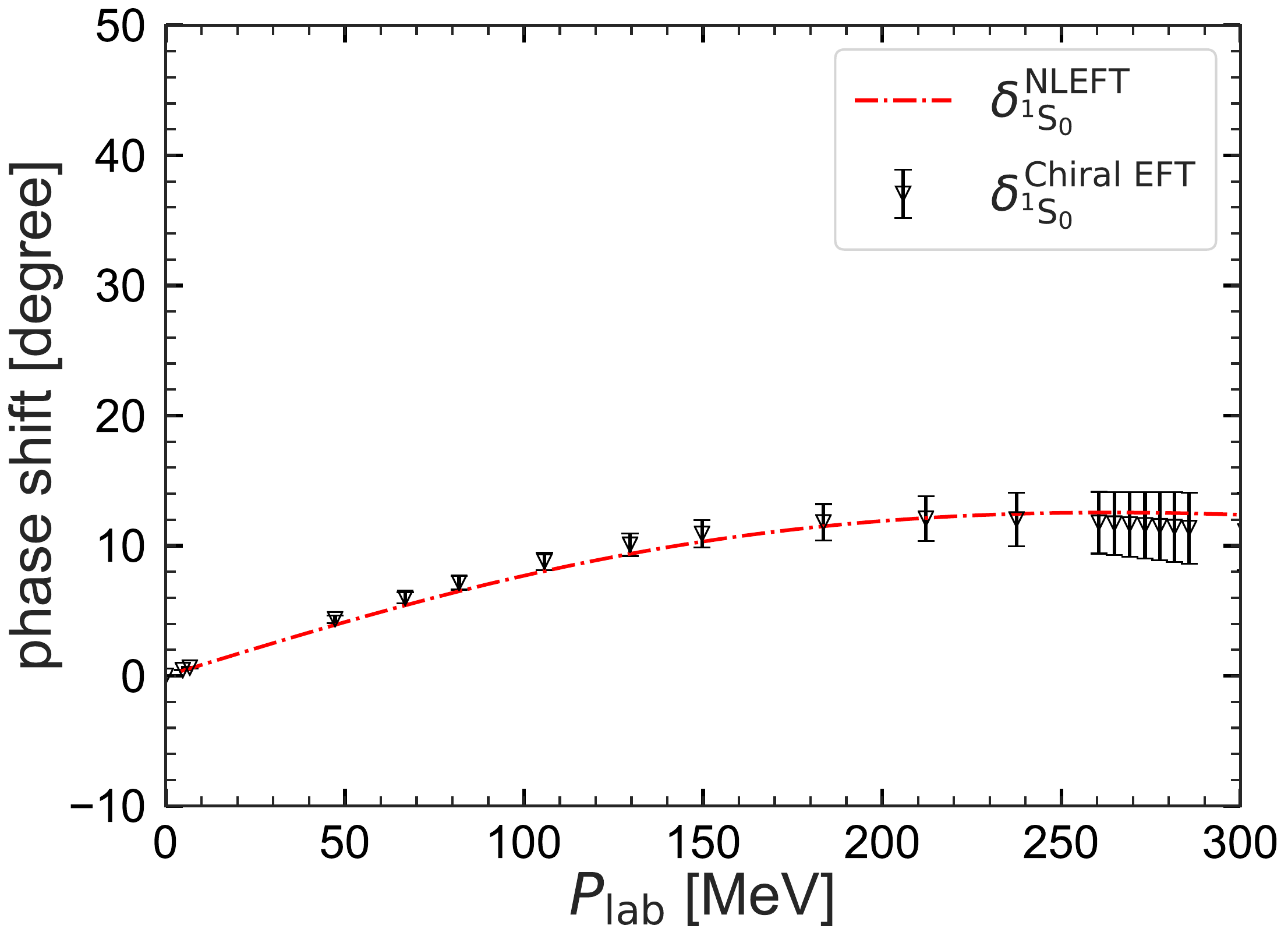}
  %\hspace{0.5cm}
  %\includegraphics[height=5.1 cm,valign=t]{figs/}
  \caption{ Baryon-baryon interactions.  Left Panel: Fit to the cross section for $N\Lambda$ scattering~\cite{Sechi-Zorn:1968mao,Alexander:1968acu,Kadyk:1971tc,Hauptman:1977hr}.
   Right Panel: Fit to the $\Lambda\Lambda$ scattering phase shift from chiral EFT~\cite{Haidenbauer:2015zqb}.
  \label{figS6}}
\end{figure}

As it is done in the nucleonic sector, to study the EoS of hyper neutron matter, we first determine the unknown LECs of the interactions involving $\Lambda$ hyperons. We start again with the two baryon-interactions.
For the $\Lambda N$ interaction, we fit experimental total cross-section data for laboratory momenta below 600~MeV, as shown
in the left panel of Fig.~\ref{figS6}. Meanwhile, for the $\Lambda\Lambda$ interaction,  given the absence of comprehensive cross-section data, we fit to the $^1{\rm S}_0$ phase shift derived from chiral EFT at next-to-leading order~\cite{Haidenbauer:2015zqb}. From these analyses, we determine the coupling constants as $c_{N\Lambda}=-6.52\times10^{-8}$~MeV$^{-2}$ and $c_{\Lambda\Lambda}=-2.96\times10^{-8}$~MeV$^{-2}$.

Similar to our nucleonic studies, we use hypernuclei to constrain the LECs of $\Lambda NN$ and $\Lambda\Lambda N$ three-baryon forces, more precisely
the ground-state energies of single-$\Lambda$ and double-$\Lambda$ hypernuclei. A direct comparison of our calculations with experimental results is given for the  separation energy, defined as
\begin{equation}
  B_\Lambda(_\Lambda^A Z)=E(^{A-1}Z)-E(_\Lambda^{A}Z),
\end{equation}
where $E$ is the energy of the system, $A$ its atomic number and $Z$ its charge. The computation of $B$ thus involves the calculation of the
energy of the nucleus $^{A-1}Z$ and the corresponding hypernucleus $_\Lambda^A Z$.
In the case of double-$\Lambda$ hypernuclei, the interesting observable we can access with the NLEFT is the double
separation energy,
\begin{equation}
  B_{\Lambda\Lambda}(_{\Lambda\Lambda}^A Z)=E(^{A-2}Z)-E(_{\Lambda\Lambda}^{A}Z)~.
\end{equation}
The calculation of this observable proceeds in the same way in the case single-$\Lambda$ hypernuclei, starting from the energy of the nucleus, the corresponding $\Lambda$ hypernucleus and now the double-$\Lambda$ hypernucleus. The best-fitting results for the separation energies of the various single-$\Lambda$ and double-$\Lambda$ hypernuclei, which are used to determine the hyperonic three-body forces given in Eqs.(\ref{eqn:V-NNY}) and (\ref{eqn:V-NYY}), are collected in Tab.~\ref{tab:hypern}.

%\begin{table}[htbp]
%    \centering
%    \caption{$\Lambda$ separation energies for single-$\Lambda$ and double-$\Lambda$ hypernuclei (in MeV). The first error is the statistical one
%    whereas the second error is the systematic one (due to the three-baryon forces). The $^*$ marks a prediction.}\label{tab:hypern}
%    \begin{tabular}{|l|c|c|}
%    \hline
%               System  & NLEFT HNM(I) HNM(II) HNM(III) & Exp.  \\
%    \hline
%    $_\Lambda^5$He   & $3.40(1)(1)$   & $3.10(3)$  \\
%    $_\Lambda^9$Be   & $5.72(5)(4)$  & $6.61(7)$   \\
%    $_\Lambda^{13}$C & $10.54(17)(29)^*$ & $11.80(16)$  \\
%    \hline
%    $_{\Lambda\Lambda}^6$He    & $7.36(1)(4)$ & $6.91(16)$  \\
%    %$_{\Lambda\Lambda}^{10}$Be & $13.30(7)(12)$ & $14.70(40)$  \\
%    %$_{\Lambda\Lambda}^{12}$Be & $21.22(56)(21)^*$ & $21.48(121)$  \\
%    \hline
%    \end{tabular}
%\end{table}

\begin{table}[htbp]
    \centering
    \caption{$\Lambda$ separation energies for single-$\Lambda$ and double-$\Lambda$ hypernuclei (in MeV). The first error is the statistical one
    whereas the second error is the systematic one (due to the three-baryon forces). The $^*$ marks a prediction.}\label{tab:hypern}
    \begin{tabular}{|c|c|c|c|c|}
    \hline
    \multirow{2}{*}{System} &
    \multicolumn{3}{c|}{NLEFT} &
    \multirow{2}{*}{Exp.}\\
    \cline{2-4}
    &HNM(I) &HNM(II) &HNM(III) &\\

    \hline
    $_\Lambda^5$He   & $3.40(1)(1)$   & $3.45(1)(2)$  & $3.46(1)(3)$ & $3.10(3)$ \\
    $_\Lambda^9$Be   & $5.72(5)(4)$  & $5.64(5)(3)$  & $5.57(5)(3)$ & $6.61(7)$\\
    $_\Lambda^{13}$C & $10.54(17)(29)^*$ & $10.09(17)(27)^*$ & $9.80(17)(26)^*$ & $11.80(16)$ \\
    \hline
    $_{\Lambda\Lambda}^6$He    & $6.91(1)(1)$ & 6.91(1)(1) & $6.91(1)(1)$ & $6.91(16)$ \\
    %$_{\Lambda\Lambda}^{10}$Be & $13.30(7)(12)$ & $14.70(40)$  \\
    %$_{\Lambda\Lambda}^{12}$Be & $21.22(56)(21)^*$ & $21.48(121)$  \\
    \hline
    \end{tabular}
\end{table}

%\textbf{
As discussed in the main text, only the separation energies of hypernuclei are utilized to constrain the $NN\Lambda$ and $N \Lambda \Lambda$ forces for HNM(I).
The hyperonic three-body forces in HNM(II) and HNM(III) are determined by
the separation energies of hypernuclei and the $\Lambda$ threshold densities $\rho_\Lambda^{\rm th}$ around $(2-3)\rho_0$.
These variations correspond to maximum neutron star masses of  $1.59(1)(1)\,M_\odot$, $1.94(1)(1)\,M_\odot$, and $2.17(1)(1)\,M_\odot$, respectively.
The HNM(III) approach, which has the largest threshold densities $\rho_\Lambda^{\rm th}$ among the three models, results in the stiffest EoS.
This finding underscores the importance of hyperonic three-body forces in achieving higher maximum neutron star masses.
%}

%references for SM
%\input{ref_SM.tex}
%\bibliography{References_SM.bib}

% \end{CJK*}
\end{document}